\documentclass[aps,prl,twocolumn,showpacs,superscriptaddress,groupedaddress]{revtex4}  % for review and submission
\usepackage{graphicx}  % needed for figures
\usepackage{dcolumn}   % needed for some tables
\usepackage{bm}        % for math
\usepackage{amssymb}   % for math

\begin{document}

% The following information is for internal review, please remove them for submission
%\leftline{Primary authors: XXX}
%\leftline{To be submitted to (PRL, PRD-RC, PRD, PLB; choose one.)}
%\leftline{Comment to {\tt d0-run2eb-nnn@fnal.gov} by xxx, yyy}
%\centerline{\em D\O\ INTERNAL DOCUMENT -- NOT FOR PUBLIC DISTRIBUTION}

% the following line is for submission, including submission to the arXiv!!
%\hspace{5.2in} \mbox{Fermilab-Pub-04/xxx-E}
\author{Benjamin B. Machta}
\affiliation{Department of Physics, Cornell University, Ithaca NY 14850}
\author{Sarah L. Veatch}
\affiliation{Department of Biophysics, University of Michigan, Ann Arbor, MI. 48109}
\author{James P. Sethna}
\affiliation{Department of Physics, Cornell University, Ithaca NY 14850}

\title{Critical Casimir forces in cellular membranes}

\begin{abstract}
Recent experiments suggest that membranes of living cells are tuned close to a miscibility critical point in the 2D Ising universality class.  We propose that one role for this proximity to criticality in live cells is to provide a conduit for relatively long-ranged critical Casimir forces.  Using techniques from conformal field theory we calculate potentials of mean force between membrane bound inclusions mediated by their local interactions with the composition order parameter.  We verify these calculations using Monte-Carlo where we also compare critical and off-critical results.  Our findings suggest that membrane bound proteins experience weak yet long range forces mediated by critical composition fluctuations in the plasma membranes of living cells.  %We argue that these results quantify one benefit cells derive by proximity to criticality.  
\end{abstract}
\pacs{87.15.kt, 87.15.Ya,87.16.dt}
\maketitle

%\section{\label{sec:level1}First-level heading}
% sections are not used for PRL papers
Cellular membranes are two-dimensional (2D) liquids composed of thousands of different lipids and membrane bound proteins.   Though once thought of as uniform solvents for embedded proteins, a wide array of biochemical and biophysical evidence suggests that cellular membranes are quite heterogeneous (reviewed in ~\cite{Pike06,Lingwood10}). Putative membrane structures, often termed `rafts', are thought to range in size from $10-100nm$, much larger than the $a \sim 1nm$ size of the individual lipids and proteins of which they are composed. This discrepancy in scale presents a thermodynamic puzzle: na\"{\i}ve estimates predict enormous energetic costs associated with maintaining heterogeneity in a fluid membrane~\cite{Machta11}. 

Parallel work in giant plasma membrane vesicles (GPMVs) isolated from living mammalian cells presents a compelling explanation for the physical basis of these proposed structures.   When cooled below a transition temperature around $25^o$C, GPMVs phase separate into two 2D liquid phases~\cite{BaumgartHSHHBW07} which can be observed by conventional fluorescence microscopy.  Quite surprisingly, they pass very near to a critical point in the Ising universality class at the transition temperature~\cite{Veatch08}.  Near a miscibility critical point, the small free energy differences between clustered and unclustered states could allow the cell to more easily control the spatial organization of the membrane, lending energetic plausibility to the proposed structures.  Although analogous critical points can be found in synthetic membranes~\cite{Veatch07,Veatch072,Honerkamp-Smith09} these systems require the careful experimental tuning of two thermodynamic parameters, as in the Ising liquid-gas transition where pressure (equivalent to the Ising magnetization) and temperature must both be tuned.  Although it has been suggested that biological systems frequently tune themselves towards \textit{dynamical} and other statistical critical points~\cite{Mora11}, so far as we know membranes are the clearest example of a biological system which appears to be tuned to the proximity of a \textit{thermal} critical point.

%Past theoretical efforts to explain `raft' heterogeneities have focused on 2D microemulsions, stabilized either by 2D surfactants~\cite{Korolev08,Brewster09} or through a coupling of composition to extrinsic curvature~\cite{Schick12}.   These theories are inspired in part by the proposed magnitude of correlations, which appear to be at odds with a fluctuation based explanation.  However, though Ising fluctuations have vanishing contrast as the domain size goes to infinity~\cite{Schick12}, in 2D, where $\beta/\nu=1/8$, spatial fluctuations remain high in contrast up to surprisingly large scales.  In a critical system, at the physiologically relevant 20nm scale, we expect fluctuations to have roughly $(20nm/1nm)^{\beta/\nu} \sim.7$ times the contrast of individual molecules. This remarkable contrast is born out experimentally in fluorescence images blurred over $400nm$~\cite{Veatch08}.   Critical fluctuations, for which there is direct experimental support in GPMVs from living cells, are sufficient to produce the high contrast heterogeneity seen in live cells.  For this reason we prefer a fluctuation based model without any additional microemulsion based mechanisms, although we expect that surfactant molecules and coupling to curvature could modulate critical parameters.

Other plausible theoretical models have focused on 2D micro emulsions (stabilized by surfactants~\cite{Brewster09}, coupling to membrane curvature~\cite{Schick12}, or topological defects in orientational order~\cite{Korolev08}) but none have emerged from direct, quantitative experiments on membranes from living cells. It has been argued that Ising fluctuations should have vanishing contrast between the two phases~\cite{Schick12}. While this is true of macroscopic regions, a region of radius $R$ of lipids of size $a\sim 1nm$ should have contrast $\sim (R/a)^{-\beta/\nu} = (R/a)^{-1/8}$, leading to predicted composition differences of $0.7$ at the physiologically relevant $20nm$ scale, and differences of 0.5 at $R=400nm$ scale of fluorescence imaging~\cite{Veatch08}; on the length scales of interest there is plenty of contrast. Indeed, our calculations of Ising-induced forces take place at and above the critical point, where the macroscopic contrast is of course zero.
 
How might a cell benefit by tuning its membrane near to criticality?   Presuming that functional outcomes are carried out by proteins embedded in the membrane, we focus on the effects that criticality might have on them.  For embedded proteins, proximity to a critical point is distinguished by the presence of large, fluctuating entropic forces known as critical Casimir forces. Three dimensional critical Casimir forces have a rich history of theoretical study\cite{Fisher78}.  In more recent experimental work~\cite{Bonn09} colloidal particles  clustered and precipitated out of suspension when the surrounding medium is brought to the vicinity of the liquid-liquid miscibility critical point in their surrounding medium.  Two dimensional Casimir forces like the ones studied here have been investigated for the Ising model using numerical transfer matrix techniques~\cite{Burkhardt95}, for a demixing transition using Monte-Carlo~\cite{Reynwar08} and for shape fluctuation using perturbative analytical methods~\cite{Yolcu11,Yolcu12}.  Here we estimate the magnitude of composition mediated Casimir forces arising from proximity to a critical point, both in Monte-Carlo simulations on a lattice Ising model, and analytically, making use of recent developments in boundary conformal field theory(CFT)~\cite{Cardy082,Ginsparg88,DiFrancesco97}.  Our motivation is biological: in a cellular membrane, these long ranged critical Casimir forces could have profound implications.  More familiar electrostatic interactions are screened over around $1nm$ in the cellular environment, whereas we find the composition mediated potential can be large over tens of nanometers.  

Critical Casimir forces are likely utilized by cells in the early steps of signal transduction where lipid mediated lateral heterogeneity has been shown to play vital roles.  Many membrane bound proteins segregate into one of two membrane phases when biochemically extracted with detergents at low temperatures~\cite{Melkonian95}, or when proteins are localized in phase separated GPMVs~\cite{Veatch08}.  Furthermore, there is evidence that some receptors change their partitioning behavior in response to ligand binding or down-stream signaling events~\cite{Holowka05}. Modeling this as a change in the coupling between the receptor protein and the Ising order parameter predicts that these bound receptors will see a change in their interaction partners.  Supporting this view, ligand binding to receptor is often accompanied by spatial reorganization in which receptors and downstream molecules move into close proximity of one another~\cite{Pike06,Veatch12}, perhaps because they now share a preference for the same Ising phase.  Perturbations to the lipid composition of the membrane, like cholesterol depletion~\cite{Levental09}, typically disrupt this spatial reorganization~\cite{Veatch12} and have dramatic effects on the final outcomes of signaling~\cite{Sheets99,Sil07,Gidwani03}, in our view by taking the membrane away from its critical point and interfering with the resulting long ranged forces.

We take three approaches to estimating the form of these potentials.  We first consider two point-like proteins which interact with the local order parameter like local insertions of magnetic field $h_1$ and $h_2$ at $x=0$ and $x=d$.  To calculate the resulting potential we write a Hamiltonian for the combined system of the Ising model with order parameter $\phi(x)$ plus proteins as $H(\left[\phi(x)\right], d) = H_{Ising}(\left[\phi(x)\right])+h_1 \phi(0)+h_2 \phi(d)$.  We then write a partition function for the combined system $Z(d)=\int D\left[\phi(x)\right]e^{-\beta H((\phi(x), d))}$ and solve to lowest order in $h$ giving the potential $U_{eff}(d) =-\log(Z(d))+\log(Z(\infty))= -h_1 h_2C(d)$ with $C(d) = \langle\phi(0)\phi(d)\rangle$ the correlation function.  $C(d) \sim d^{-\eta}$ when $d\ll \xi$ with the Ising model $\eta=\frac{1}{4}$ and $C(d) \sim d^{-1/2}\exp(-d/\xi)$ for $d\gg \xi$.  The potential is attractive for like and repulsive for unlike insertions of field, in agreement with the scaling of the CFT result as we will show below.   A protein which does not couple to the order parameter can still feel a long-ranged force if it couples to the local energy density.  The energy density is also correlated with a $d^{-2}$ dependance.  However, the magnitude of both of these potentials, as well as their shape at distances $d\sim r$  require the Monte-Carlo and CFT approaches described below.

Secondly, we numerically calculated potentials using Monte-Carlo on the lattice Ising model for like and unlike disk-shaped inclusions.  Although absolute free energies are difficult to obtain from Monte-Carlo techniques, differences between the free energies of two ensembles, $\delta F$, conditioned on a subset of the degrees of freedom are readily available, provided the degrees of freedom in the two `macro-states' can be mapped onto each other and have substantial overlap.  This information is implicitly used in a Monte-Carlo scheme where both `macro-states' are treated as members of a larger ensemble and are switched between so as to satisfy detailed balance. The Bennett method~\cite{Bennett76,Jarzynski97}, uses this information more explicitly, noting that $\exp{(-\beta\delta F)}=\left\langle e^{-\beta \delta E}\right\rangle$  can be estimated without bias from either distribution.

\begin{figure}
\includegraphics[scale=.5]{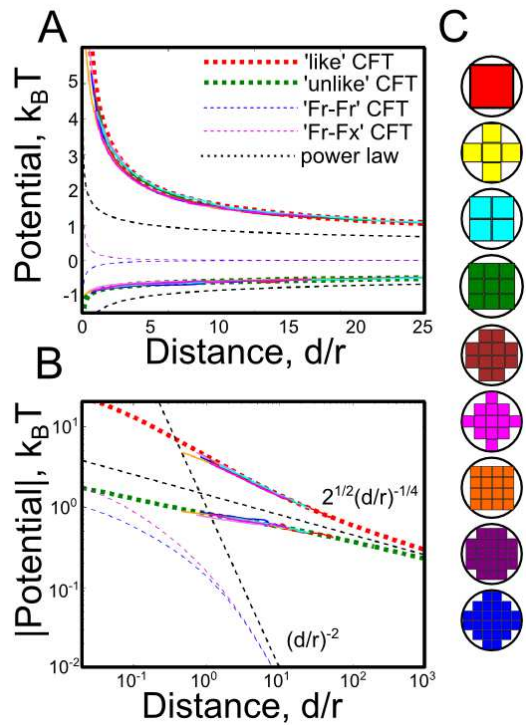}
\caption{\label{fig:Feff} Effective potentials between bound inclusions are plotted on linear (top) and log-log (bottom) graphs, for inclusions where $r_1=r_2=r$.  The CFT results for both like and unlike interactions (thick dashed lines) and for potentials containing a free BC agree with the power-law scaling of the two-point function (thin black dashed line) at large lengths, but separate at small separations.  We also compare to Bennett method simulations at $T_c$ as described in the text.  We run simulations for each of the blocky spheres shown in (C).  Each curve is plotted collapsed by using $r$ as the distance to the farthest point from its center, with no free parameters.  The results of our Monte-Carlo pair potentials are all shown plotted against $d/r$ (thin solid lines with colors as in (C)) with the theory curves in dashed lines.  The CFT prediction is in excellent agreement with simulation data even for very small inclusions well past the applicability of the power law prediction of the perturbative approach.  The value of the potential is fit at the farthest accessible simulation point, where we add the CFT prediction.}
\end{figure}

Our `macro-states' are the location of two blocky `disks' as shown in fig~\ref{fig:Feff}C.  All spins either contained in or sharing a bond with these disks are constrained to be either all up or all down.  We map the degrees of freedom in one macro-state to a neighboring one by moving all of the spin values 1 lattice spacing to the right or left of the fixed spin region onto fixed spins on the other side. By integrating our measured $\beta \delta F=-\log \left<exp(-\beta \delta E )\right>  $ over many sites outwards to infinity, we can in principle measure this potential to arbitrary distance.  However, because the potential is long-ranged at $T_c$, we integrate it out to $50$ lattice spacings and add the CFT prediction for the potential at that distance as described below.  We perform simulations using the Wolff Algorithm on $500 \times 500$ lattices under the constraint that any cluster which intersects a disk is rejected, enforcing our fixed boundary conditions.  We supplement these with individual spin flips near the inclusions where almost all Wolff moves are rejected.  The resulting potentials are plotted in fig.~\ref{fig:Feff}A.  We collapse the Monte-Carlo curves by using the the effective radius given by the farthest point from the origin contained in the blocky lattice inclusion as the effective radius.

\begin{figure}
	\includegraphics[scale=0.45]{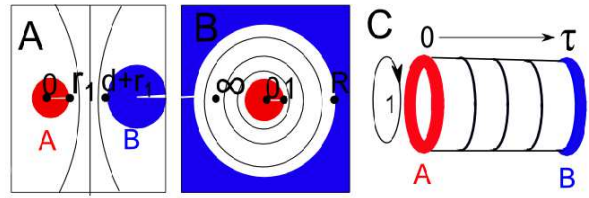}
	\caption{\label{fig:ConformalMapping} We consider potentials of mean force in configuration (A), with disks of radius $r_1$ and $r_2$ separated by a distance $d$ with boundary conditions $A$ and $B$. We conformally map this to configuration (B), where both disks are centered on the origin, with the first at radius $1$ and the second at radius $R(d,r_1,r_2)$.  We then map this to a cylinder shown in (C) of circumference $1$ and length $-i\tau=\log(R)/2\pi$ where we associate restricted partition functions in an imaginary time $1+1$D quantum model with potentials of mean force in the original configuration.}
\end{figure}

Finally, we use conformal field theory to make an analytical prediction for the form of these potentials.  Our calculation makes extensive use of the conformal invariance of the free energy which emerges at the critical point.   An element from the global conformal group can take us from the configuration in fig.~\ref{fig:ConformalMapping}A to that shown in fig.~\ref{fig:ConformalMapping}B where the two disks are concentric with spatial infinity in fig.~\ref{fig:ConformalMapping}A now lying between the two cylinders on the real axis. The radius of the outer circle $R(d,r_1,r_2)$ is now given by:
\begin{equation}
\label{eq:Rdef}
\begin{array}{clrr}
R(d,r_1,r_2)=\frac{x-2+\sqrt{(x-2)^2-4}}{2} \text{, } x=\frac{(d+2r_1)(d+2r_2)}{r_1r_2}
\end{array}
\end{equation}

The much larger local conformal group, particular to 2D, is the set of all analytic functions.  We use the transformation $z' =\frac{\log(z)}{2\pi}$ gluing together the boundaries at $x=1$ and $x=0$ to give the cylinder shown in fig.~\ref{fig:ConformalMapping}C with a circumference of $1$ and length: 
\begin{equation}
\label{eq:taudef}
	\tau(d,r_1,r_2)=i\log(R(d,r_1,r_2))/2\pi
\end{equation}	
This transformation breaks global conformal invariance and so increases the free energy by $c\log(R)/12$~\cite{Ginsparg88}, where $c=1/2$ in the Ising model.  Defining a $1+1$ dimensional quantum theory on the cylinder (see ~\cite{Ginsparg88}) with `time', $t$ running down its length, our Hamiltonian for $t$ translation is $H=2\pi(L_0+\bar{L}_0-\frac{c}{12})$, where $L_0+\bar{L}_0$ is the generator of dilation in the plane.

Partition functions in this geometry are linear sums of characters of the conformal group.  The representations of the conformal group particular to the Ising universality class have characters given by~\cite{Ginsparg88,Cardy08}:
\begin{equation}
\label{eq:chidef}
\begin{array}{clrr}
\chi_{0}(\tau)=\frac{1+q^2+q^3+\cdots}{q^{1/48}}=\frac{1}{2\sqrt{\eta(\tau)}}\left[\sqrt{\theta_3(q)}+\sqrt{\theta_4(q)}\right] \\
\chi_{1/16}(\tau)=\frac{1+q+q^2+2q^3+\cdots}{q^{1/48-1/16}}=\frac{1}{\sqrt{2\eta(\tau)}}\left[\sqrt{\theta_2(q)}\right] \\
\chi_{1/2}(\tau)=\frac{1+q+q^2+\cdots}{q^{1/48-1/2}}=\frac{1}{2\sqrt{\eta(\tau)}}\left[\sqrt{\theta_3(q)}-\sqrt{\theta_4(q)}\right]
\end{array} 
\end{equation}
where $q=\exp{(i\pi\tau)}$, with $\eta(\tau)$ the Dedekind $\eta$ function and with $\theta(\tau)$ the Jacobi, or elliptic Theta functions.
 
Conformally invariant boundary conditions (BCs) can be deduced by demanding consistency between two parameterizations of the cylinder~\cite{Cardy08}.  In one, time moves from one BC to the other across the cylinder with the usual Ising Hamiltonian.  Alternatively,  time can move around the cylinder with the BCs now entering into the Hamiltonian. There are three allowed BCs~\cite{Cardy08} which, by considering symmetry can be associated with `up', `down' and `free'. These three BCs have four non-trivial potentials between them; a repulsive `unlike' interaction between `up' and `down' BCs, an attractive `like' interaction between `ups' and `ups' or `downs' and `downs', an attractive `free-free' (Fr-Fr) interaction between two `free' BCs and a repulsive `free-fixed' (Fr-Fx) interaction between a `free' BC and either an `up' or a `down'.  

 The free energy in the configuration shown in figure ~\ref{fig:ConformalMapping}A can be interpreted as a potential of mean force between the bound inclusions.  Choosing the convention that the potentials go to 0 as $d \rightarrow \infty$, the potential is given by $U(d)= F_{AB}(\tau)-F_{AB}(\infty)$.  After undoing the mapping which changes the free energy by a central charge dependent factor so that $F_{AB}(\tau)=- \log{Z_{AB}(\tau)} + c\pi\tau/6$ (with $k_BT=1$) the potentials are given by: 
\begin{equation}
\label{eq:potential}
\begin{array}{llll}
U_{\text{like}}(d,r_1,r_2)\\
\text{ }=-\log \left(\chi_{o}(2\tau)+\chi_{1/2}(2\tau)+\sqrt{2}\chi_{1/16}(2\tau)\right)+\frac{\pi\tau}{12}\\
U_{\text{unlike}}(d,r_1,r_2)\\
\text{ }=-\log \left(\chi_{o}(2\tau)+\chi_{1/2}(2\tau)-\sqrt{2}\chi_{1/16}(2\tau)\right)+\frac{\pi\tau}{12}\\
U_{\text{Fr-Fr}}(d,r_1,r_2)=-\log \left(\chi_{o}(2\tau)+\chi_{1/2}(2\tau)\right)+\frac{\pi\tau}{12}\\
U_{\text{Fr-Fx}}(d,r_1,r_2)=-\log \left(\chi_{o}(2\tau)-\chi_{1/2}(2\tau)\right)+\frac{\pi\tau}{12}\\
\end{array}
\end{equation}
with $\chi_h$ as defined in eq.~\ref{eq:chidef}, and $\tau$ as defined in eqs.~\ref{eq:Rdef} and~\ref{eq:taudef}.  These potentials are plotted on regular and log-log graphs in figure~\ref{fig:Feff}.  Their form is in agreement with the numerical results obtained using transfer matrix methods in~\cite{Burkhardt95}.

At large $d$, we can examine the asymptotics of the potentials using the form of each potential in eq.~\ref{eq:potential} and the series expansion of the characters as shown in eq.~\ref{eq:chidef}. For fixed BCs, the leading contribution to the potential of mean force is equal to $\pm \sqrt{2(r_1r_2)^{1/4}}d^{-\frac{1}{4}}$, with a sign which differs depending on whether the two BCs are like or unlike, in agreement with the point like approximation.  For potentials that involve at least one `free' BC, similar analysis shows that the leading contribution is proportional to $d^{-2}$.  All four potentials diverge at short distances like $\pm d^{-1/2}$ where in all cases the sign is positive unless both BCs are identical.   We note that the origins of the two techniques leading to the curves shown in fig.~\ref{fig:Feff} are very different; arguably as different from each other as each are from a lipid bilayer.  The very close agreement, even at lengths comparable to the lattice spacing speaks to the power of universality.

\begin{figure}
	\includegraphics[scale=0.5]{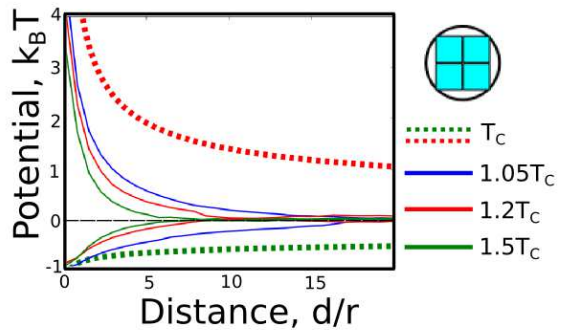}
	\caption{\label{fig:vsT} We compare our critical results with potentials obtained from Monte-Carlo simulations away from the critical point along the temperature axis.  As can be seen, the potentials are longest ranged at the critical point.  The repulsive interaction is also steepest at the critical point, though the attractive one has a larger force at short distances slightly away from the critical point. }
\end{figure}

 We also compare the form of the potential with Monte-Carlo results performed at temperatures away from the critical point where the potential has a range given roughly by $\xi$.  In each case the resulting potential is a one dimensional cut through a four dimensional scaling function which could depend nontrivially on $d/r_1$,$d/r_2$,$d/\xi$ and the `polar' coordinate $h/t^{\beta\delta}$~\cite{Schofield69} describing the proximity to criticality. The dashed lines show the CFT prediction for $T=T_c$, with numerical results at $1.05$,$1.1$ and $1.2T_c$, all for the $2 \times 2$ block sphere shown at right in fig.~\ref{fig:vsT}.  The repulsive potential is both deepest and sharpest at $T_c$, while the the attractive force is sharpest slightly above $T_c$, with the final potential of very similar magnitude.

We expect our results to apply, with a few caveats, to proteins embedded in real cell membranes.  Proteins couple to their surrounding composition through the height of their hydrophobic regions, interactions of their membrane-proximal amino acids with their closest lipid shell and by covalent attachment to certain lipids which themselves strongly segregate into one of the two low temperature phases.  In simulation our proteins couple strongly to their nearest neighbor lipids leading to potentials in excellent agreement with CFT predictions that are very different in origin.  These are expected to describe any uniform boundary condition in an Ising liquid, in the limit where all lengths are large compared to the lattice spacing.  When separated by lengths of order a lipid spacing (1nm) we might expect additional corrections to this form, and in particular, a weakly coupled protein may have behavior intermediate between a `free' and a `fixed' BC.  In addition, a protein that couples non-uniformly around its boundary might have interesting behavior not addressed here.   We note that our boundary conditions couple to two long-ranged scaling fields- the magnetization field which falls off with the a power of $-1/4$ and the energy density which falls off with a power of $-2$, both of which must be present in membranes or any other system near an Ising critical point. 

It is interesting to compare this composition mediated force to other forces that could act between membrane bound proteins.  Electrostatic interactions are screened over around $1nm$ in the cellular environment, making them essentially a contact interaction from the perspective of the cell.  There is an analogous shape fluctuation mediated Casimir force that falls off like $d^{-6}$~\cite{Yolcu11,Yolcu12}, and is therefore also very short ranged.  Membrane curvature can also mediate forces with a leading attractive term that falls off like $d^{-2}$ and a leading repulsive term that falls off like $d^{-4}$.  Although they decay with a much larger power than the critical Casimir forces described above, curvature mediated potentials depend on elastic constants and are not bound to be of order $k_BT$ allowing them to become quite large at shorter distances.  Using typical values~\cite{Reynwar07} the potentials are comparable at lengths $\sim 5-10nm$ to the composition mediated potential we find here~\cite{Dommersnes99}.  There are numerous examples of biology using these relatively short ranged but many $k_BT$ potentials for coordinating energetically expensive and highly irreversible events like vesiculation~\cite{Reynwar07}.  We propose that critical Casimir forces could mediate long ranged and reversible interactions useful for regulating a protein's binding partners. More generally, this work demonstrates that the hypothesis of criticality enables a quantitative understanding of the broad range of phenomena frequently associated with `raft' heterogeneity in cell membrane.

%(XXX Sarah What do you think?)Widespread evidence suggests that cells make use of these long ranged forces, especially in signal transduction. Signaling events often involve spatial reorganization at the membrane surface, with receptors and downstream molecules moving into close proximity of each other (reviewed in~\cite{Pike06}).  This reorganization~\cite{xxxEthan'sPaper?}, as well as functional outcomes of the signaling cascade~\cite{Sheets99}, can be disrupted by a large class of lipid perturbations, including cholesterol depletion and loading, and the addition of short chain ceramides~\cite{Gidwani03}.  These perturbations may act as a change in the Ising parameters of reduced temperature and magnetization which describe the proximity to criticality~\cite{xxx?}.  In addition, after signaling molecules have been added, some receptors purify into different membrane fractions~\cite{xxxHolowkaRef?}, suggesting that ligand binding has changed their preference for the local lipid composition.  Finally, experiments using structurally defined multi-valent ligands connected with variable length DNA linkers show that functional outcomes are sensitive to the geometric positioning of receptors at lengths $\sim 10nm$~\cite{Sil07}, suggesting that some relatively long ranged force is being used to communicate between them.

This work was supported by NIH R00GM087810, NSF DMR 1005479, and NIH T32GM008267. We thank Paul Ginsparg, Chris Henley, Markus Deserno, Cem Yolcu, Barbara Baird and David Holowka for useful discussions.

\bibliographystyle{prsty}
\bibliography{membraneref095}

\begin{thebibliography}{10}

\bibitem{Pike06}
L.~J. Pike, J Lipid Res {\bf 47},  1597  (2006).

\bibitem{Lingwood10}
D. Lingwood and K. Simons, Science {\bf 327},  46  (2010).

\bibitem{Machta11}
B. Machta, S. Papanikolaou, J. Sethna, and S. Veatch, Biophysical Journal {\bf
  100},  1668  (2011).

\bibitem{BaumgartHSHHBW07}
T. Baumgart {\it et~al.}, Proc. Natl. Acad. Sci. {\bf 104},  3165  (2007).

\bibitem{Veatch08}
S.~L. Veatch {\it et~al.}, ACS Chemical Biology {\bf 3},  287  (2008).

\bibitem{Veatch07}
S.~L. Veatch, Seminars in Cell \& Developmental Biology {\bf 18},  573  (2007).

\bibitem{Veatch072}
S.~L. Veatch, O. Soubias, S.~L. Keller, and K. Gawrisch, Proc. Natl. Acad. Sci.
  USA {\bf 104},  17650  (2007).

\bibitem{Honerkamp-Smith09}
A.~R. Honerkamp-Smith, S.~L. Veatch, and S.~L. Keller, Biochimica et Biophysica
  Acta - Biomembranes {\bf 1788},  53  (2009).

\bibitem{Mora11}
T. {Mora} and W. {Bialek}, Journal of Statistical Physics  135  (2011).

\bibitem{Brewster09}
R. Brewster, P.~A. Pincus, and S.~A. Safran, Biophysical Journal {\bf 97},
  1087  (2009).

\bibitem{Schick12}
M. Schick, Phys. Rev. E {\bf 85},  031902  (2012).

\bibitem{Korolev08}
K.~S. Korolev and D.~R. Nelson, Phys. Rev. E {\bf 77},  051702  (2008).

\bibitem{Fisher78}
M. Fisher and P. Gennes, Comptes Rendus Hebdomadeires Des Seances De L Academie
  Des Sciences Serie B {\bf {287}},  207  (1978).

\bibitem{Bonn09}
D. Bonn {\it et~al.}, Physical Review Letters {\bf {103}},    ({2009}).

\bibitem{Burkhardt95}
T.~W. Burkhardt and E. Eisenriegler, Phys. Rev. Lett. {\bf 74},  3189  (1995).

\bibitem{Reynwar08}
B.~J. Reynwar and M. Deserno, Biointerphases {\bf 3},  FA117  (2008).

\bibitem{Yolcu11}
C. Yolcu, I.~Z. Rothstein, and M. Deserno, {EPL} {\bf {96}},    ({2011}).

\bibitem{Yolcu12}
C. Yolcu, I.~Z. Rothstein, and M. Deserno, Phys. Rev. E {\bf 85},  011140
  (2012).

\bibitem{Cardy082}
J. Cardy, Boundary Conformal Field Theory, 2008.

\bibitem{Ginsparg88}
P. {Ginsparg}, {Applied Conformal Field Theory, arXiv:hep-th/9108028}  (1991).

\bibitem{DiFrancesco97}
P. Di~Francesco, P. Matthieu, and P. S$\acute{e}$n$\acute{e}$chal, {\em
  Conformal Field Theory} (Springer-Verlag, New York, NY, 1997).

\bibitem{Melkonian95}
K.~A. Melkonian, T. Chu, L.~B. Tortorella, and D.~A. Brown, Biochemistry {\bf
  34},  16161  (1995).

\bibitem{Holowka05}
D. Holowka {\it et~al.}, Biochimica et Biophysica Acta - Molecular Cell
  Research {\bf 1746},  252  (2005).

\bibitem{Veatch12}
S.~L. Veatch {\it et~al.}, The Journal of Physical Chemistry B, In print
  (2012).

\bibitem{Levental09}
I. Levental {\it et~al.}, Biochemical Journal {\bf {424}},  163  (2009).

\bibitem{Sheets99}
E. Sheets, D. Holowka, and B. Baird, Journal of Cell Biology {\bf {145}},  877
  (1999).

\bibitem{Sil07}
D. Sil {\it et~al.}, ACS Chemical Biology {\bf 2},  674  (2007).

\bibitem{Gidwani03}
A. Gidwani, H. Brown, D. Holowka, and B. Baird, Journal of Cell Science {\bf
  116},  3177  (2003).

\bibitem{Bennett76}
C. Bennett, Journal Of Computational Physics {\bf {22}},  245  ({1976}).

\bibitem{Jarzynski97}
C. Jarzynski, Physical Review Letters {\bf 78},  2690  (1997).

\bibitem{Cardy08}
J. Cardy, Conformal Field Theory and Statistical Mechanics, 2008.

\bibitem{Schofield69}
P. Schofield, Physical Review Letters {\bf {22}},  606  ({1969}).

\bibitem{Reynwar07}
B.~J. Reynwar {\it et~al.}, Nature {\bf 447},  461  (2007).

\bibitem{Dommersnes99}
P. Dommersnes and J.-B. Fournier, The European Physical Journal B - Condensed
  Matter and Complex Systems {\bf 12},  9  (1999).

\end{thebibliography}

\end{document}